\documentclass[pra,aps,showpacs,twocolumn,superscriptaddress]{revtex4-1}
\usepackage{amsmath,amssymb,graphicx}
\usepackage{amsmath}
\usepackage{bm}
\usepackage{braket}
\usepackage{float}
\usepackage[11pt]{moresize}
\usepackage[caption=false]{subfig}
\usepackage{mathtools}

\begin{document}
	
\title{Why a hole is like a beam splitter\---  \\
	a general diffraction theory for multimode quantum states of light}	

\author{Zhihao Xiao}
\email[zxiao3@lsu.edu]{}
\author{R. Nicholas Lanning}
\affiliation{Hearne Institute for Theoretical Physics, and Department of Physics $\&$ Astronomy, Louisiana State
	University, Baton Rouge, Louisiana 70803, USA}

\author{Mi Zhang}
\author{Irina Novikova}
\author{Eugeniy E. Mikhailov}
\affiliation{Department of Physics, College of William $\&$ Mary, Williamsburg, Virginia 23187, USA}

\author{Jonathan P. Dowling}
\affiliation{Hearne Institute for Theoretical Physics, and Department of Physics $\&$ Astronomy, Louisiana State
	University, Baton Rouge, Louisiana 70803, USA}

\begin{abstract}
Within the second-quantization framework, we develop a formalism for describing a spatially multimode optical field diffracted through a spatial mask and show that this process can be described as an effective interaction between various spatial modes. We demonstrate a method to calculate the quantum state in the diffracted optical field for any given quantum state in the incident field. Using numerical simulations,  we also show that with single-mode squeezed-vacuum state input, the prediction of our theory is in qualitative agreement with our experimental data. We also give several additional examples of how the theory works, for various quantum input states, which may be easily tested in the lab; including two single-mode squeezed vacuums, single- and two-photon inputs, where we show the diffraction process produces two-mode squeezed vacuum, number-path entanglement and a Hong-Ou-Mandel-like effect\---analogous to a beam splitter.
\end{abstract}	
	
\maketitle 

\section{Introduction}
Gaussian spatial modes, in comparison with plane waves, offer a more accurate description of optical beams \cite{Siegman_book}. Although plane waves are mathematically simpler, they are less powerful in describing the diffraction and the spatial structure of optical fields. Classical diffraction properties of Gaussian beams are relatively well understood, and numerous works have been carried out, both in theory and experiment \cite{zhang1989spatial,liu1982simple,belland1982changes,pearson1969diffraction,uehara1989generation,porras1998ultrashort,oemrawsingh2004production,passilly2005simple}.  The quantum properties of diffracted Gaussian beams, or other paraxial beams, have received less attention, although some of previous works are found in Refs. \cite{pevrinova2006quantization,treps2002surpassing,treps2003quantum,lassen2009continuous,banaszek2001generation,corzo2011multi,janousek2009optical,beck2000quantum,la1991squeezing,kim1994quadrature,chalopin2011direct,lopez2009multimode,chalopin2011multimode,kolobov1999spatial,kolobov2003sub,chalopin2010multimode,vaziri2002superpositions}. Though many previous analyses do take multiple Gaussian modes under consideration, a clear and systematic description of the interaction amongst Gaussian modes is lacking. By Gaussian-mode interaction we mean all physical processes in which the output spatial mode decomposition is altered from its input decomposition. The assumption, taken in many cases, that Gaussian modes interact in the same way as plane waves, is generally not valid because it essentially ignores the multi-mode structure of Gaussian modes. 
Gaussian modes are a natural choice to describe propagation of optical beams with finite cross section. 
Indeed, if the squeezed states generated in different Gaussian modes have different squeezing angles, then the interaction among the states in the various modes can worsen rather than improving the overall squeezing. Recent work further confirms the deficiency of using plane waves to analyze quantum states of light, and motivates us to investigate the quantum behavior of Gaussian beams \cite{zhang2016spatial,zhang2013generating,horrom2012quantum}. 

To understand the quantum behavior of Gaussian beams, we must understand how quantum states in different Gaussian modes interact with each other. 
Perhaps the simplest interaction between Gaussian modes can be introduced by applying a spatial mask on the beam axis and seeing how this would change the quantum states. Through this relatively simple model, we can establish a method to analyze more complicated problems.

In Sec. II, we use classical electrodynamics to analyze the Gaussian beam and the interactions among different orders of Gaussian modes. In Sec. III, we present the quantum description of states in Gaussian beams and their interactions. In Sec. IV, we present our result from numerical simulations, which agrees with our experimental data. In Sect. V, we will consider three examples of applying our formalism to describe propagation of various quantum input optical fields through an iris mask. The first one uses two single-mode squeezed vacuums as input, which has been tested experimentally \cite{zhang2016spatial}, and our predictions agree well with the experimental observations. The other examples study the cases of a single photon, and two-photon inputs, in which case our calculations predict the generation of a photon-number entanglement and a Hong-Ou-Mandel-like effect,  implying that an opaque spatial mask displays characteristics of a regular but lossy optical beam splitter.




\section{Classical Electrodynamic description of Gaussian beam spatial modes}
%
%
%
%
%
%
%
%
%
%

\begin{figure}
	[htbp]
	\centering
	\includegraphics[width=1.0\columnwidth]{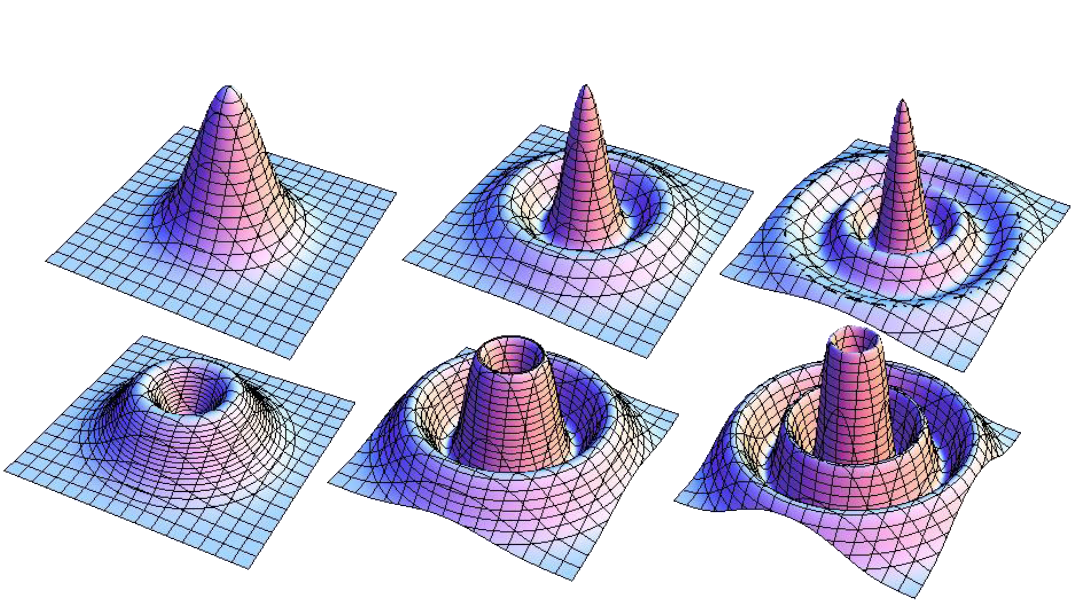}
	\caption{\label{fig:LG_intensity_profile_6_fancy} 
		Intensity profile of LG modes in any $z=z_0$ plane. Upper row (from left to right):  $l=0, p=0,1,2$; lower row: $l=1,p=0,1,2$. }
\end{figure}

For an optical beam, the electromagnetic field satisfies Maxwell's equations in the so called paraxial approximation. Furthermore, it is known that the Hermite-Gaussian (HG) and Laguerre-Gaussian (LG) modes are solutions of free-space wave equation in the paraxial approximation. In Cartesian coordinates the solutions are the HG modes whereas in cylindrical coordinates the solutions are the LG modes. While we focus on the LG modes for the rest of this paper, similar arguments apply to the HG modes. The normalized field amplitude of LG modes can be expressed as follows:

\begin{equation} \begin{split}
	{u}_{ l,p}(r,\phi,z)=\frac{C^{\rm{LG}}_{lp}}{w(z)}\left(\frac{r \sqrt{2}}{w(z)}\right)^{|l|}\exp\left(-\frac{r^2}{w^2(z)}\right)\\
	L_p^{|l|} \left(\frac{2r^2}{w^2(z)}\right) 	\exp\left( - i k \frac{r^2}{2 R(z)}\right)\\
	\exp(i l \phi)\exp\left[i(2p+|l|+1)\zeta(z)\right],
\end{split}\end{equation}
where $r$, $\phi$ and $z$ are cylindrical coordinates; $l$ and $p$ are the azimuthal and radial indices, which are integers; $p \geqslant 0$; $C^{\rm{LG}}_{lp}=\sqrt{\frac{2}{\pi} \frac{p!}{(|l|+p)!}}$ is a normalization constant; 
$L_p^{|l|}$ is the associated Laguerre polynomial; $k$ is the wave number; 
$w(z)=w_0 \sqrt{1+(\frac{z}{z_R})^2}$ is the beam waist; $w_0$ is the beam waist at the beam focus;  $z_R=\frac{\pi w_0^2}{\lambda}$ is the Rayleigh range; $R(z)=z[1+(\frac{z_R}{z})^2]$ is the radius of curvature; $\zeta(z)=\arctan (\frac{z}{z_R})$ is the Gouy phase. See Fig.~\ref{fig:LG_intensity_profile_6_fancy} for the intensity profile of several LG modes in any $z=z_0$ plane. Along the beam axis the profile will become wider or narrower with changes of the beam waist, while the shapes of the profiles remain similar.

In free space the LG modes propagate independently without interacting with each other; they obey the following orthonormality conditions,

\begin{equation}\label{eqn:orthogonality}
	\int_{z=z_0} u_{ l,p} u_{ l',p'}^* r dr d\phi  = \delta_{ ll'} \delta_{ pp'} .
\end{equation}
Notice, the orthogonality condition only holds if the integration area on the left hand side of Eq.~(\ref{eqn:orthogonality}) is the entire $z\!=\!z_0$ plane. 

\begin{figure}
	[htbp]
	\centering
	\includegraphics[width=1.0\columnwidth]{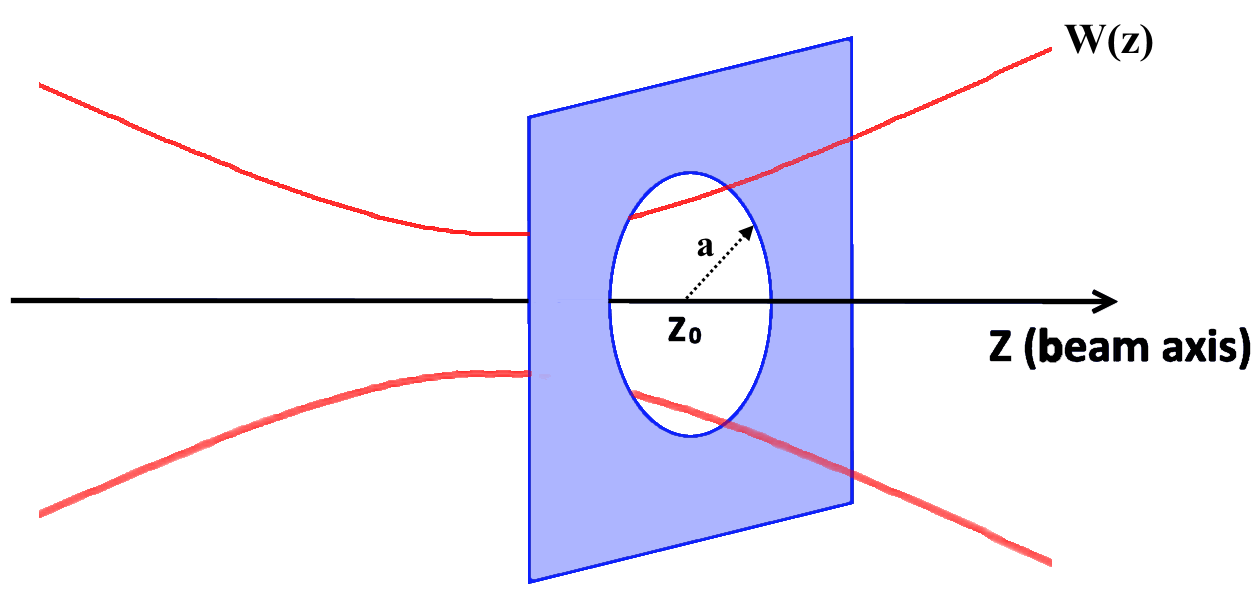}
	\caption{\label{fig:waist_iris_axis_radius_font_black_w}
		The iris, with a circular opening of radius $a$, is applied along the beam axis. The red curve is the Gaussian beam width $w(z)$ as a function of $z$. The iris is located at the plane $z=z_0$. The center of iris is on the $z$ axis. The amplitude will be truncated to zero at the rim of the iris, while in the opening of the iris the amplitude will be unchanged. As a result the orthogonality between LG modes will be broken, and the modes will interact at the iris plane.}
\end{figure}

Now, let us consider putting a spatial mask (such as a circular iris) in the $z=z_0$ plane, shown in Fig.~\ref{fig:waist_iris_axis_radius_font_black_w}. The iris blocks or absorbs the field at the rim and allows the field at the opening to pass through. For LG modes, the part of them allowed to pass through the opening of iris no longer obeys orthogonality. Physically this means different LG modes will interact at the plane where the iris is placed. The interaction of modes can be described by the following expression,

\begin{equation}\label{eqn:u_to_b}
\int\displaylimits_{S} u_{ l,p} u_{ l',p'}^* r dr d\phi  = B_{l,l',p,p'},
\end{equation}
where $S$ is the surface through which the spatial mask permits the light to pass. For a circular iris with radius $a$, centered on the beam axis, placed in $z=z_0$ plane, $S=\{r<a; z=z_0\}$.

Since in free space LG modes form an orthonormal basis, both the input signal (at $z={z_0}^-$) and the output signal (at $z={z_0}^+$) can be expressed as linear combinations of LG modes, which both satisfy paraxial approximation. 
Further, in free space on both sides of the iris [$z \in (- \infty , z_0) \cup (z_0, + \infty) $], the orthogonality of LG modes holds, and the iris ($z=z_0$) is the only location where orthogonality breaks. Therefore the coefficient of each LG mode will change only when the signal goes through the iris. We express this interaction using the following set of equations: the input beam takes the form,
\begin{equation}\label{eqn:input}
	{u}_{\textrm{input}}(r,\phi,z)=\sum_{l,p}A_{l,p}\times{u}_{ l,p}(r,\phi,z);\ (z<z_0),
\end{equation}
where $A_{l,p}$ is the coefficient of each LG mode. At the iris the beam is partially absorbed and thus we have,
\begin{equation}\label{eqn:iris}
	{u}_{\textrm{iris}}(r,\phi,z_0)=
	\begin{cases}
		\sum_{l,p}A_{l,p}\times{u}_{ l,p}(r,\phi,z_0);\ (r<a,z=z_0) \\
		\\
		0;\ (r\geq a,z=z_0), \\
	\end{cases}
\end{equation}
satisfying the boundary condition at the iris, giving the output signal,  
\begin{equation}\label{eqn:boundary}
	{u}_{\textrm{iris}}(r,\phi,z_0)={u}_{\textrm{output}}(r,\phi,{z_0}^+) ,
\end{equation}
which finally leads to,
\begin{equation}\label{eqn:output}
{u}_{\textrm{output}}(r,\phi,z)=\sum_{l,p}A_{l,p}\sum_{l',p'}B_{l,l',p,p'}\times{u}_{ l',p'}(r,\phi,z) ;\  (z>z_0).
\end{equation}
The quantity $B_{l,l',p,p'}$, first introduced in Eq.~(\ref{eqn:u_to_b}), is the transformation coefficient between LG mode $l, p$ and $l', p'$. Solving Eqs.(\ref{eqn:input}, \ref{eqn:iris}, \ref{eqn:boundary}, \ref{eqn:output}), we get,
\begin{equation}\label{eqn:B}
	B_{l,l',p,p'}=C_{l,l',p,p'} \times \exp[i(2p-2p'+|l|-|l'|)\zeta(z_0)].
\end{equation}
Here we express the complex quantity $B_{l,l',p,p'}$ in polar form, as it more clearly shows the role of $\zeta(z_0)$, which is the Gouy phase at the iris position. The factor $C_{l,l',p,p'}$ is real in the circular iris situation, since the cylindrical symmetry prevents interaction between azimuthal indexes:
\begin{equation}\label{eqn:C}
C_{l,l',p,p'}=\delta_{ ll'} \times \int\limits_{0}^{2a^2/w^2(z_0)}\exp{(-x)} L_p^{|l|}(x) L_{p'}^{|l'|}(x) dx.
\end{equation}

It is due to the limitation in the radial direction, introduced by the iris, that different $p$ modes will interact. But due to the cylindrical symmetry of the iris, different $l$ modes will remain orthogonal. Therefore, if instead of a circular iris, other types of spatial masks (that do not have cylindrical symmetry) are used, orthogonality amongst $l$ modes will be broken, and different $l$ modes will interact with each other. The interactions, which are characterized by the transformation coefficient, will be determined by the shape and position of the spatial mask. For the remainder of this paper, we consider only an iris spatial mask because of its simplicity. However our theory applies to all spatial masks, and the transformation coefficients can be calculated in a similar manner. To calculate transformation coefficients for an arbitrary spatial mask, we can still make use of the more general Eq.~(\ref{eqn:u_to_b}) instead of Eqs.~(\ref{eqn:B}, \ref{eqn:C}), which are specifically suitable for circular iris mask centered on beam axis.

\section{Quantization of Gaussian modes}
%

In free space, due to orthogonality, each mode of the Gaussian beam propagates without interacting with another. Therefore, the quantum state of each mode will evolve independently. A pure quantum state without mode entanglement is a product state of every quantum state in every Gaussian mode:

\begin{equation}\label{eqn:separable_state}
	\ket{\psi}=\prod_{l=-\infty, p=0}^{l=+\infty, p=+\infty} \ket{\psi_{l,p}}.
\end{equation}

The separable state forms a building block for more complicated states. A general pure state, with or without mode entanglement, can be expressed as a linear combination of separable states in the form of Eq.~(\ref{eqn:separable_state}). Further, a mixed state can be expressed as probabilistic sum of pure states.

\begin{figure}
	[htbp]
	\centering
	\includegraphics[width=1.0\columnwidth]{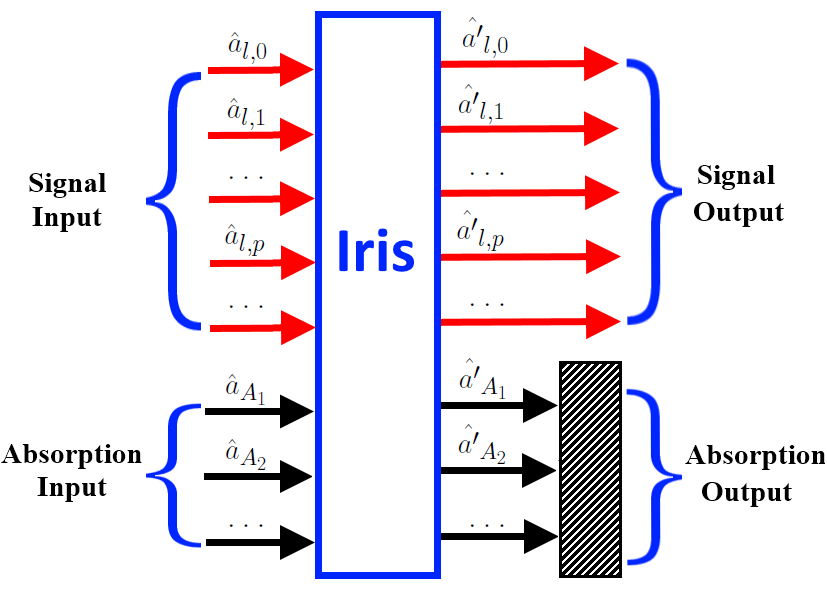}
	\caption{\label{fig:iris_diagram_super_shortened_3_w}
		The iris transforms the creation and annihilation operators of the input modes into operators of the output modes. Since the rim of the iris blocks off part of the beam, absorption modes (denoted as $A_1, A_2$ etc.), in addition to the original Gaussian beam modes (signal modes), are needed. The input states in the absorption modes are vacuum. To obtain the reduced density operator in the output-signal modes, the states in the output absorption modes need to be traced out.}
\end{figure}

When a spatial mask such as an iris is applied to the Gaussian beam, the quantum states of different modes will interact. The interaction can be described as a transformation of annihilation or creation operators of input modes into operators of output modes. This transformation should be unitary, which preserves the commutation relations of the annihilation or creation operators. However, one problem needs to be solved. Generally, spatial masks (or other optical devices) are lossy. For example, an iris will absorb part of the input signal at the rim. A widely accepted procedure \cite{leonhardt1997measuring} to deal with loss in quantum optics is to introduce ``absorption modes'' which we denote as: $A_1,A_2,...$. To be clear, we call the original Gaussian modes ``signal modes'', since they are the ones that may contain information, such as squeezing levels and squeezing angles.  We denote the signal modes with simply $l$ and $p$ numbers. Further, we denote with a prime symbol on the operators of output modes to differentiate them from the input modes. The transformation, caused by the iris, from the operators of input modes into those of output modes, is illustrated in Fig.~\ref{fig:iris_diagram_super_shortened_3_w}.

Before we continue, let us explain a bit more about the absorption modes. They serve three purposes. The first purpose is that they describe the absorption (loss) of the field. Since the states in output-signal modes are described by tracing the entire output density matrix over the absorption modes, the total energy of the signal is generally decreased. The second purpose is that they help to keep the transformation unitary by expanding the dimension of the transformation matrix \cite{Gerry_and_Knight_book}. The reader might remember a similar principle applies when modeling loss with a simple beam splitter; we must consider a second input even if only the first input is used \cite{leonhardt1997measuring}. The third purpose of the absorption modes is that they naturally introduce the vacuum fluctuations and accommodate the common observation that fluctuations usually occur with losses.

The model works in the following way. The quantum states of the input signal modes can be arbitrary, but quantum states in the input-absorption modes are vacuum. A unitary transformation transforms the operators of input-signal or absorption modes into operators of output-signal or absorption modes; illustrated in Fig.~\ref{fig:iris_diagram_super_shortened_3_w}. Once we obtain output operators in terms of input operators we can then calculate the quantum states in the output-signal or absorption modes. The quantum state in all output modes then needs to be traced over the output absorption modes, and finally we obtain the reduced density matrix that describes the quantum state in the output-signal modes, which generally is a mixed state, even if the input state is a pure state.  Later we will give a few examples for a variety of input states.

\subsection{Matrix description of Gaussian mode interactions}

The interaction among the quantum states of all LG modes can be described with a transformation from the operators of input modes (signal and absorption) to the operators of output modes. Such a transformation, as previously argued, is unitary for spatial masks. There are infinitely many Gaussian modes (and we need to introduce infinitely many absorption modes as well). Therefore, in the most general case, quantum states or operators in infinitely many input modes are transformed into quantum state or operators in infinitely many output modes. 

Although this might seem complicated, sometimes the transformation can be greatly simplified when the spatial mask has some kind of symmetry. For example, as we previously pointed out, an iris has cylindrical symmetry and LG modes with different $l$'s do not interact (due to the Kronecker delta in Eq.~(\ref{eqn:C}), which enforces angular momentum conservation). Therefore, for a circular iris, we need only to examine the transformation of LG modes with the same $l$ but different $p$'s. To that end, we introduce the column vector of annihilation operators for input LG mode $(l,p=0), (l,p=1), (l,p=2)$ etc. as well as operators for input-absorption modes $A_1, A_2, A_3$ etc, 

\begin{equation}\label{eqn:column_a_l}
	(\hat{a_l}) =\begin{pmatrix} \hat{a}_{l,0} & \hat{a}_{l,1} & \cdots & \hat{a}_{A_1} & \hat{a}_{A_2} & \cdots \end{pmatrix}^T.
\end{equation}
The creation operators are similarly,
\begin{equation}
(\hat{a_l}^{\dagger}) =\begin{pmatrix} \hat{a}^{\dagger}_{l,0} & \hat{a}^{\dagger}_{l,1} & \cdots & \hat{a}^{\dagger}_{A_1} & \hat{a}^{\dagger}_{A_2} & \cdots \end{pmatrix}^T,
\end{equation}
and the output modes follow, but they are marked with prime
\begin{equation}
(\hat{a'_l}) =\begin{pmatrix} \hat{a'}_{l,0} & \hat{a'}_{l,1} & \cdots & \hat{a'}_{A_1} & \hat{a'}_{A_2} & \cdots \end{pmatrix}^T,
\end{equation}
\begin{equation}
(\hat{a'_l}^{\dagger}) =\begin{pmatrix} \hat{a'}^{\dagger}_{l,0} & \hat{a'}^{\dagger}_{l,1} & \cdots & \hat{a'}^{\dagger}_{A_1} & \hat{a'}^{\dagger}_{A_2} & \cdots \end{pmatrix}^T.
\end{equation}
We also define the unitary transformation matrix $\mathcal{J}_l$, which determines the interaction among LG modes with same $l$, but different values of $p$.
\begin{equation}
\mathcal{J}_l
=
\begin{bmatrix}
J_{l;0,0}       & J_{l;0,1} &   \dots & J_{l;0,A_1} & J_{l;0,A_2} &  \dots \\
J_{l;1,0}       & J_{l;1,1} &   \dots & J_{l;1,A_1} & J_{l;1,A_2} &  \dots \\
\hdotsfor{6} \\
J_{l;A_1,0}       & J_{l;A_1,1} &   \dots & J_{l;A_1,A_1} & J_{l;A_1,A_2} &  \dots \\
J_{l;A_2,0}       & J_{l;A_2,1} &   \dots & J_{l;A_2,A_1} & J_{l;A_2,A_2} &  \dots \\
\hdotsfor{6} \\
\end{bmatrix}.
\end{equation}


The transformation of input and output operators can be expressed in the following compact form
\begin{equation}
\label{eqn:short_matrix_transformation}
(\hat{a_l}')=\mathcal{J}_l (\hat{a_l}),
\end{equation}
\begin{equation}
\label{eqn:short_matrix_transformation_dagger}
(\hat{a_l'}^{\dagger})=\mathcal{J}_l^* (\hat{a_l}^{\dagger}).
\end{equation}
$\mathcal{J}_l^*$ stands for the conjugate (without transpose) of $\mathcal{J}_l$. The signal-signal elements ($J_{l;0,0},J_{l;1,0},J_{l;0,1}$, etc.) in the matrix $\mathcal{J}_l$ determine the transformation between input and output-signal modes. Here we make use of Bohr's correspondence principle. For large amplitude coherent states in the input signal modes, the transformation between input and output-signal modes should agree with the classical result in Eq.~(\ref{eqn:output}), giving, 
\begin{equation}
J_{l;p_1,p_2}=B_{l,l,p_1,p_2},
\end{equation} 
which can be calculated using Eqs.~(\ref{eqn:B}, \ref{eqn:C}). As for the other (signal-absorption and absorption-absorption) elements in $\mathcal{J}_l$, we can make use of $\mathcal{J}_l$ being unitary. This gives $\mathcal{J}_l \mathcal{J}_l^{\dagger}=I$, which will give equations describing the relations among the $\mathcal{J}_l$ elements. Of course, signal-absorption and absorption-absorption elements might not be completely fixed, and there might be a certain freedom of choice. In fact, they may not need to be calculated at all. We find that, in the calculations we have done so far, we can always eliminate signal-absorption and absorption-absorption elements using the condition that $\mathcal{J}_l$ is unitary.

Indeed, if one aims for completeness, one should consider infinitely many LG modes. However, we do not usually have that luxury, since the dimension of the transformation matrix grows with the number of modes, and we are forced to consider a limited number of modes. Intuitively, the more modes we consider, the better. But the effect of higher-order modes often diminishes at a very fast rate. As we show in the next section, we are able to explain our experimental data, even if we consider only two input signal modes and two absorption modes. 

For a spatial mask with arbitrary shape, we cannot exploit the cylindrical symmetry as we did with the iris. However, we can still introduce similar column vector of operators as before, but we now will need to include various $l$ modes together instead of only considering one $l$ mode at a time. We can achieve this by defining a concatenation of column vectors of operators, such as $(\hat{a}) =\begin{pmatrix} (\hat{a}_{l=0})^T & (\hat{a}_{l=1})^T & (\hat{a}_{l=-1})^T & \cdots \end{pmatrix}^T$, in which every element is defined in Eq.~(\ref{eqn:column_a_l}). The transformation matrix $\mathcal{J}$ between input to output modes needs to be expanded in similar fashion in order to accommodate different $l$ modes; and the integration area of Eq.~(\ref{eqn:u_to_b}) needs to be changed as well. Then we can finally arrive at the relation similar to Eq.~(\ref{eqn:short_matrix_transformation}): $(\hat{a}')=\mathcal{J} (\hat{a})$. 

Unlike the iris, a spatial mask without cylindrical symmetry introduces interaction between orbital angular momentum modes, which can be very useful. However the purpose of this work is not to explore novel designs of optical devices, but to setup a general method for analyzing a range of problems. For now, the simple iris is enough to serve such a purpose, but we stress that our method can also accommodate optical devices without cylindrical symmetry.

\section{Numerical Simulation of our Experiment}
Let us consider the following model of our recent experiment \cite{zhang2016spatial}. In the two signal LG modes of ($l=0, p=0$) and ($l=0, p=1$), we input two squeezed-vacuum quantum states, which are defined as $\hat{S}(\xi_0) \ket{0}_{l=0, p=0}$ and $\hat{S}(\xi_1) \ket{0}_{l=0, p=1}$, respectively, while in every other LG mode we input vaccum states. The squeezing operators are defined as $\hat{S}(\xi_p)=\exp [\frac{1}{2}(\xi_p^* \hat{a}_{0,p}^2 -\xi_p \hat{a}_{0,p}^{\dagger 2} )]$, with $p=0, 1$. The squeezing parameters are $\xi_0=r_0\exp(i \theta_0)$ and $\xi_1=r_1\exp(i \theta_1)$.

Let us further consider a classical field with large amplitude, in the ($l=0, p=0$) LG mode, acting as a local oscillator (LO) for homodyne detection. The signal and local oscillator co-propagate with each other along the beam axis, but they are in perpendicular polarizations.

\begin{figure}
	[htbp]
	\centering
	\includegraphics[width=1.0\columnwidth]{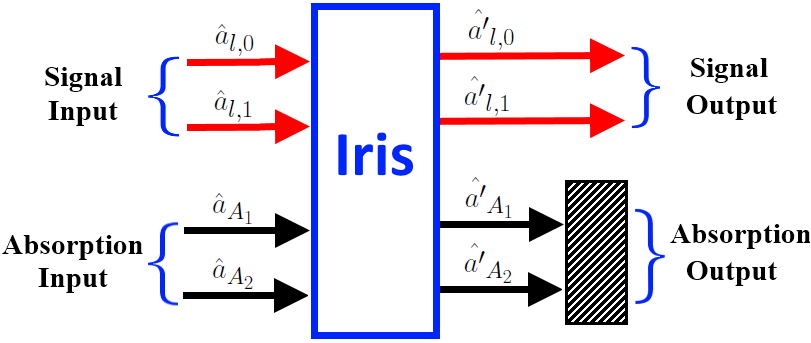}
	\caption{\label{fig:iris_diagram_shortened_twomodes_short_iris_20160806}
		Instead of taking into account of infinitely many input or output modes, we consider only two input- (or output-) signal modes and two absorption modes, because the input states in the ($l=0, p=0$) and ($l=0, p=1$) LG modes are the only ones that are non-vacuum, and the interaction between the two modes far exceeds the interaction between other LG modes. }
\end{figure}

Now we insert a circular iris in the neighborhood of the beam focus point and centered on the beam axis. According to our theory, both the signal and LO are influenced by the iris in the way described in previous sections. Introduced by the iris, the interaction among LG modes mainly happens between ($l=0, p=0$) and ($l=0, p=1$) modes. Therefore we can simplify the calculation by considering only two input- (or output-) signal modes and two absorption modes, instead of taking into account infinitely many input (or output) modes. The diagram of this model is shown in Fig.~\ref{fig:iris_diagram_shortened_twomodes_short_iris_20160806}.  We then move the iris along the beam axis, and numerically simulate the minimum noise measured in the homodyne detection vs.\! the iris position, shown in Fig.~\ref{fig:MinNoise_vs_IrisPosition_0,3_0,4_0π4_1,3π4}. We can also use different-sized irises, which are represented by different colored curves. 

\begin{figure}
	[htbp]
	\centering
	\includegraphics[width=1.0\columnwidth]{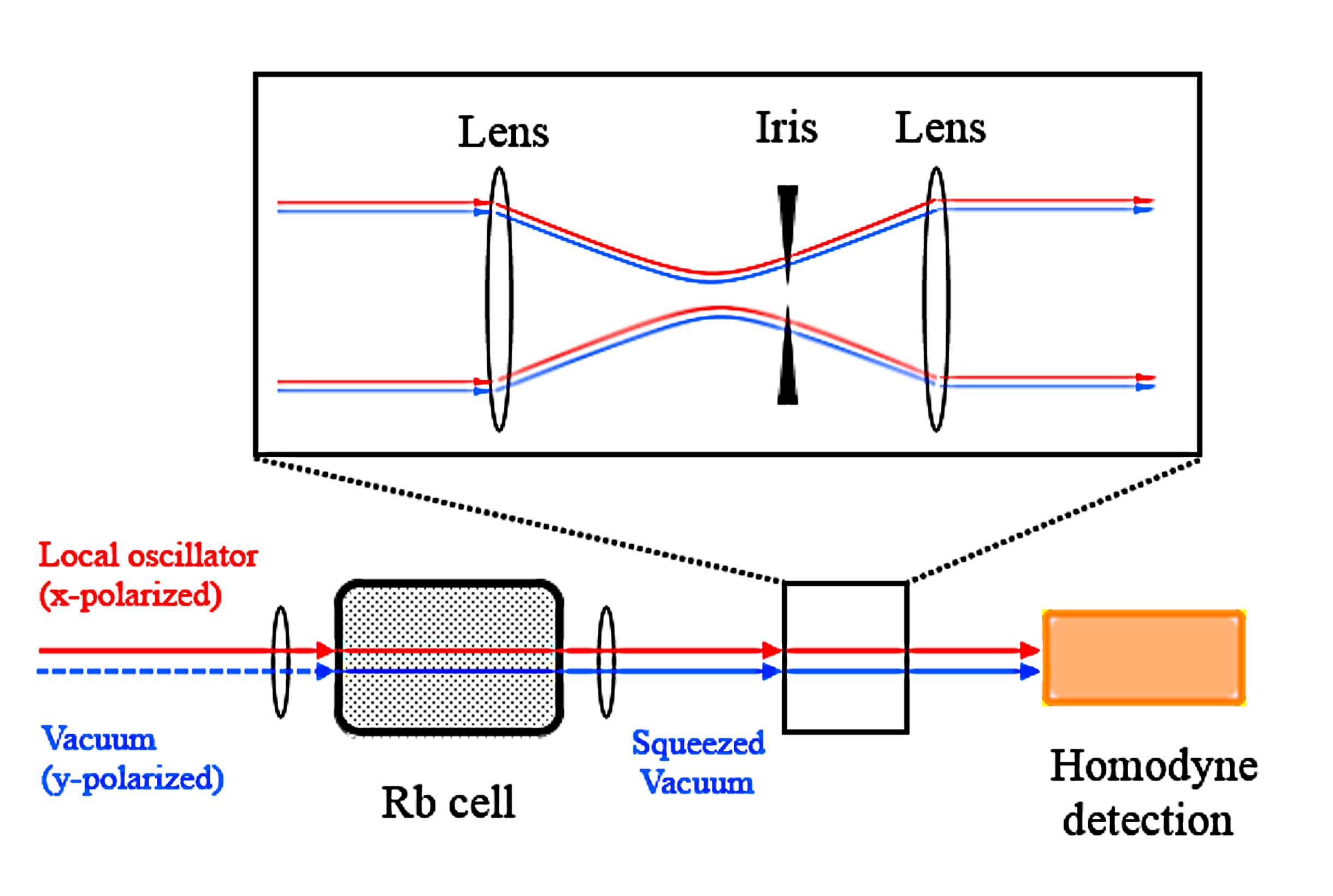}
	\caption{\label{fig:experimental_setup_redrawn_1_w}
		Experimental setup. The spatial mask consists of a one-to-one telescope and an iris between the lenses. We move the iris along the beam axis and find the minimum noise in each case. }
\end{figure}

%
%
\begin{figure}
	\centering
	\subfloat[\label{fig:MinNoise_vs_IrisPosition_0,3_0,4_0π4_1,3π4}]{\includegraphics[width=1.0\columnwidth]{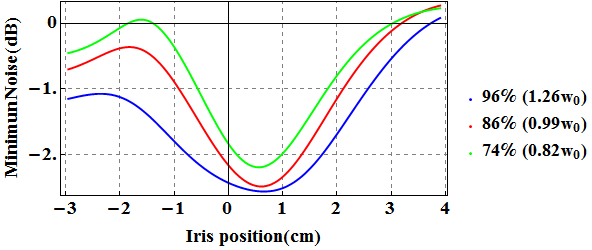}}
	\hspace{0mm}
	\subfloat[\label{fig:ExperimentPlot_MinNoise_vs_IrisPosition_replot}]{\includegraphics[width=1.0\columnwidth]{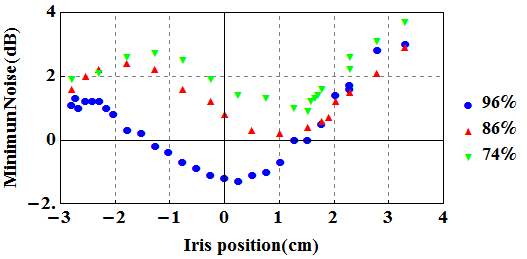}}
	\caption{(a) Numerical simulation: noise measured in homodyne detection vs the iris position. The Rayleigh range $z_R=2.5$cm. Different sized irises are represented by different curve colors, and they are denoted by the percentage of peak transmission through the iris relative to full beam transmission, as well as the iris radius (scaled by $w_0$). We only apply one iris at a time. The input states in the ($l=0, p=0$) and ($l=0, p=1$) LG modes are squeezed states with different squeezing parameters: $r_0=0.3$, $\theta_0=0$, $r_1=0.4$, $\theta_1=0.325\pi$. (b) Experimental plot: noise measured in homodyne detection vs. the iris position. The Rayleigh range $z_R=2.5$cm. Each symbol is denoted by the percentage of peak transmission through the iris relative to full beam transmission, hence represents a different iris size. The zero coordinate is the position of the beam focus and positive direction is to the right of it.}
\end{figure}

To verify our theory, we now realize this model in our experiment \cite{zhang2016spatial}, which has the following setup, shown in Fig.~\ref{fig:experimental_setup_redrawn_1_w}.
A strong, $x$-polarized beam, originally in ($l=0, p=0$) LG mode, is focused into a Pyrex cell filled with Rb atoms. 
The $y$-polarized field is coherent vacuum. 
The nonlinear interaction between the atoms and the strong field generates a squeezed-vacuum field in the $y$-polarized direction, which is in a multi-mode structure, and it is the lower two $p$ modes that contribute the most to the squeezing, and states in nonzero $l$ modes are vacuum because of conservation of angular momentum \cite{zhang2016spatial}.
Amplified by the strong field, the noise in this squeezed field will be then detected in a homodyne detection scheme. 
An iris is introduced in the neighborhood of the beam focus point and centered on the beam axis. 
A plot of minimum noise in the signal field, vs. the iris location along the axis, has been made in our previous work \cite{zhang2016spatial} (see Fig.~\ref{fig:ExperimentPlot_MinNoise_vs_IrisPosition_replot}).

Comparing our numerical simulation (Fig.~\ref{fig:MinNoise_vs_IrisPosition_0,3_0,4_0π4_1,3π4}) and our experimental data (Fig.~\ref{fig:ExperimentPlot_MinNoise_vs_IrisPosition_replot}), we find that they qualitatively agree with each other. In our experiment, thermal noise brings up the noise level, which explains the noise-level (vertical) shift between the experimental plot and the numerical simulation. A more general situation for squeezed state input in two LG modes is discussed in Section V(A).

\section{Additional Examples of the use of the theory}

\subsection{Example 1: Squeezed-vacuum input states and the Wigner function description}

We now model inputs of displaced, and non-displaced, single-mode squeezed-vacuum states in the two signal modes: ($l=0, p=0$) and ($l=0, p=1$) LG modes. For simplicity we ignore the other higher-order LG modes. In addition we include two absorption modes $A_1, A_2$. The two input-signal states have different squeezing levels and squeezing angles. We then put an iris at various locations along the beam axis and calculate the reduced density matrix of the output state in each of the two signal modes. 

In this example, we express the quantum state using the Wigner function, since it offers a clearer and more intuitive view. Now the question becomes: how does the iris transform the input Wigner function into the output Wigner function? To address this question we first use Eq.~(\ref{eqn:short_matrix_transformation}) and (\ref{eqn:short_matrix_transformation_dagger}) to calculate the transformation between input and output modes operators, and obtain the transformation of quadratures,

\begin{equation}\label{eqn:transformQ}\begin{split}
\begin{bmatrix}
q_{l,0} \\
q_{l,1} \\
q_{A_1} \\
q_{A_2} \\
\end{bmatrix}
=
\text{Re}(\mathcal{J}_l)\begin{bmatrix}
q_{l,0}' \\
q_{l,1}' \\
q_{A_1}' \\
q_{A_2}' \\
\end{bmatrix}-\text{Im}(\mathcal{J}_l)\begin{bmatrix}
p_{l,0}' \\
p_{l,1}' \\
p_{A_1}' \\
p_{A_2}' \\
\end{bmatrix},
\end{split}\end{equation}

\begin{equation}\label{eqn:transformP}\begin{split}
\begin{bmatrix}
p_{l,0} \\
p_{l,1} \\
p_{A_1} \\
p_{A_2} \\
\end{bmatrix}
=
\text{Re}(\mathcal{J}_l)\begin{bmatrix}
p_{l,0}' \\
p_{l,1}' \\
p_{A_1}' \\
p_{A_2}' \\
\end{bmatrix}+\text{Im}(\mathcal{J}_l)\begin{bmatrix}
q_{l,0}' \\
q_{l,1}' \\
q_{A_1}' \\
q_{A_2}' \\
\end{bmatrix}.
\end{split}\end{equation}
Then we substitute the input quadratures with output quadratures, thus completing the transformation of input Wigner function to output Wigner function.

\begin{equation}\begin{split}
W(q_{0,0},p_{0,0},q_{0,1},p_{0,1},q_{A1},p_{A1},q_{A2},p_{A2})\\
\xrightarrow[]{\text{Eq.~(\ref{eqn:transformQ})(\ref{eqn:transformP})}}W(q_{0,0}',p_{0,0}',q_{0,1}',p_{0,1}',q_{A1}',p_{A1}',q_{A2}',p_{A2}')
\end{split}\end{equation}

In this example the input states of the iris are displaced squeezed states: $\hat{D}(\alpha_0) \hat{S}(\xi_0) \ket{0}_{l=0, p=0}$ and $\hat{D}(\alpha_1) \hat{S}(\xi_1) \ket{0}_{l=0, p=1}$ in ($l=0, p=0$) and ($l=0, p=1$) LG modes. The displacement operator for $l=0, p=1,2$ mode is defined as $\hat{D}(\xi_p)=\exp (\alpha_p \hat{a}_{0,p}^{\dagger} -\alpha_p^* \hat{a}_{0,p})$. The Wigner functions of the quantum states in the two input-signal modes are:

\begin{equation}\begin{split}
&W(q_m,p_m)\\
&=\frac{1}{\pi}\exp\bigl\{-e^{-2r_m}[(p_m-\bar{p}_m)\cos{(\theta_m/2)}\\
&-(q_m-\bar{q}_m)\sin{(\theta_m/2)}]^2\bigr\}\\
&\times\exp\bigl\{-e^{2r_m}[(q_m-\bar{q}_m)\cos{(\theta_m/2)}\\
&+(p_m-\bar{p}_m)\sin{(\theta_m/2)}]^2\bigr\},
\end{split}\end{equation}
where $q_m, p_m$ are the quadratures of mode ($l=0, p=m$) and $m=0, 1$, and $\bar{q}_m=\frac{1}{\sqrt{2}}(\alpha_m+\alpha_m^*),  \bar{p}_m=\frac{i}{\sqrt{2}}(-\alpha_m+\alpha_m^*)$, $\xi_m=r_m\exp(i \theta_m)$. For absorption modes the input states are vacuum, whose Wigner functions are:

\begin{equation}\begin{split}
W(q_n,p_n)=\frac{1}{\pi}\exp(-q_n^2-p_n^2),
\end{split}\end{equation}
where $n=A_1, A_2$. Since the total input state is a product state of states in each of the four input modes, the total input Wigner function is:

\begin{equation}\begin{split}
&W(q_{0,0},p_{0,0},q_{0,1},p_{0,1},q_{A1},p_{A1},q_{A2},p_{A2})\\
&=W(q_{0,0},p_{0,0})W(q_{0,1},p_{0,1})W(q_{A1},p_{A1})W(q_{A2},p_{A2}).
\end{split}\end{equation}

\begin{figure}
	\centering
	\includegraphics[width=1.0\columnwidth]{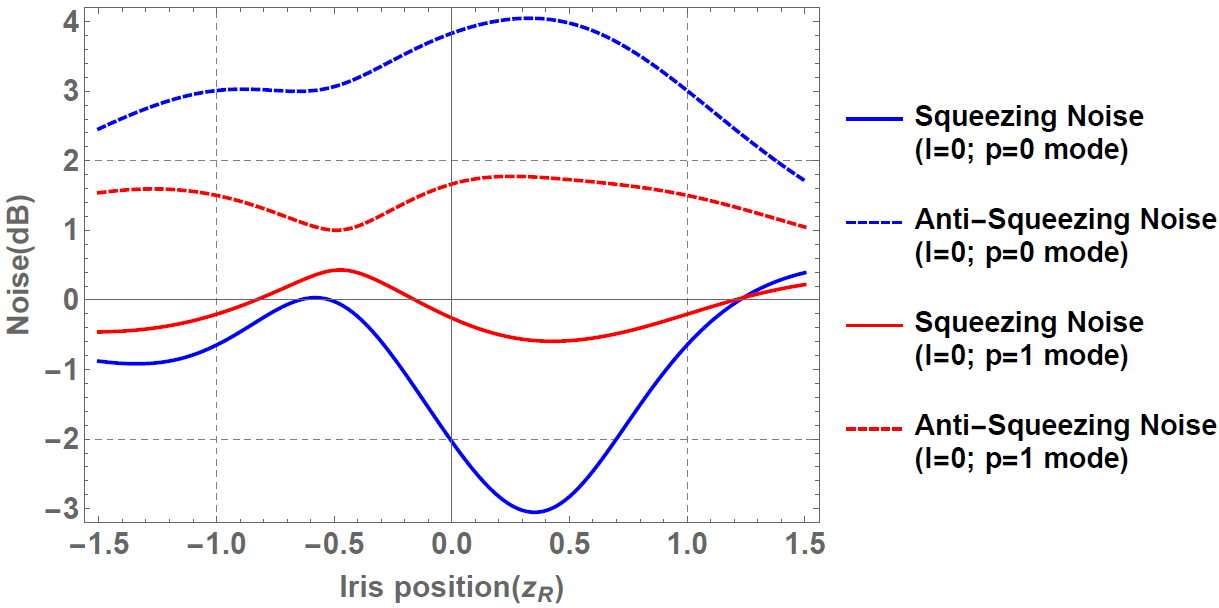}
	\caption{\label{fig:Noise_vs_iris_posi_two_modes_red&blue&dashed}
		Noise of squeezing and anti-squeezing of $l=0, p=0$ and $l=0, p=1$ LG modes vs. iris position, which is shown in the units of Rayleigh range. The parameters of the input states and iris size are the same with the parameters described in the caption of Fig.~\ref{fig:double_column_Wigner_1,0w0_narrow_contour}. Notice that the squeezing and anti-squeezing noise curve for of the same mode are not symmetrical about the horizontal axis ($0$dB noise level line) for displaced-squeezed input states. This is because the quantum state in each LG mode is no longer a minimum uncertainty state. 
	}
\end{figure}

We keep the input state fixed and change the position of the iris along beam axis. The change in iris position changes the matrix elements of $\mathcal{J}_{l=0}$ in Eqs.~(\ref{eqn:transformQ}, \ref{eqn:transformP}), which in turn changes the output states in three ways: (a) the squeezing and anti-squeezing level changes, (b) the squeezing angle changes, and (c) the state displacement (from vacuum) changes. The Wigner functions of the input state (as well as output states) of different iris positions are plotted in Fig.~\ref{fig:double_column_Wigner_1,0w0_narrow_contour}. It is also worth noting that, despite the input-signal states in this example being pure states (displaced squeezed-vacuum states in each of the two LG modes), the output-signal states are generally mixed states. This is mainly because we obtain the reduced density operator for the signal modes by tracing the total density operator over the absorption modes. As a result, the output states are no longer pure minimum uncertainty states. To verify this we can simulate the  squeezing and anti-squeezing noise in each output LG mode vs. the iris position; shown in Fig.~\ref{fig:Noise_vs_iris_posi_two_modes_red&blue&dashed}.  For a minimum-uncertainty squeezed state, the squeezing and anti-squeezing noise should add up to $0$ dB, which means the squeezing and anti-squeezing noise curve for of the same mode should be symmetric about the horizontal axis in Fig.~\ref{fig:Noise_vs_iris_posi_two_modes_red&blue&dashed}. This is obviously not the case, which verifies that the output-signal state is not a minimum uncertainty state in either LG mode. The noise measurement in Fig.~\ref{fig:Noise_vs_iris_posi_two_modes_red&blue&dashed} is achievable in an experiment. We would need the local oscillator in one particular single output LG mode to be measured with homodyne detection, which can be done with a spatial light modulator. Notice the noise measurement described in Fig.~\ref{fig:MinNoise_vs_IrisPosition_0,3_0,4_0π4_1,3π4} is different. In that previous case, the local oscillator co-propagates with the signal and both of them are influenced by the iris; after the iris the local oscillator consists of multiple LG modes instead of a single mode. 

\begin{figure*}
	\includegraphics[width=2.0\columnwidth]{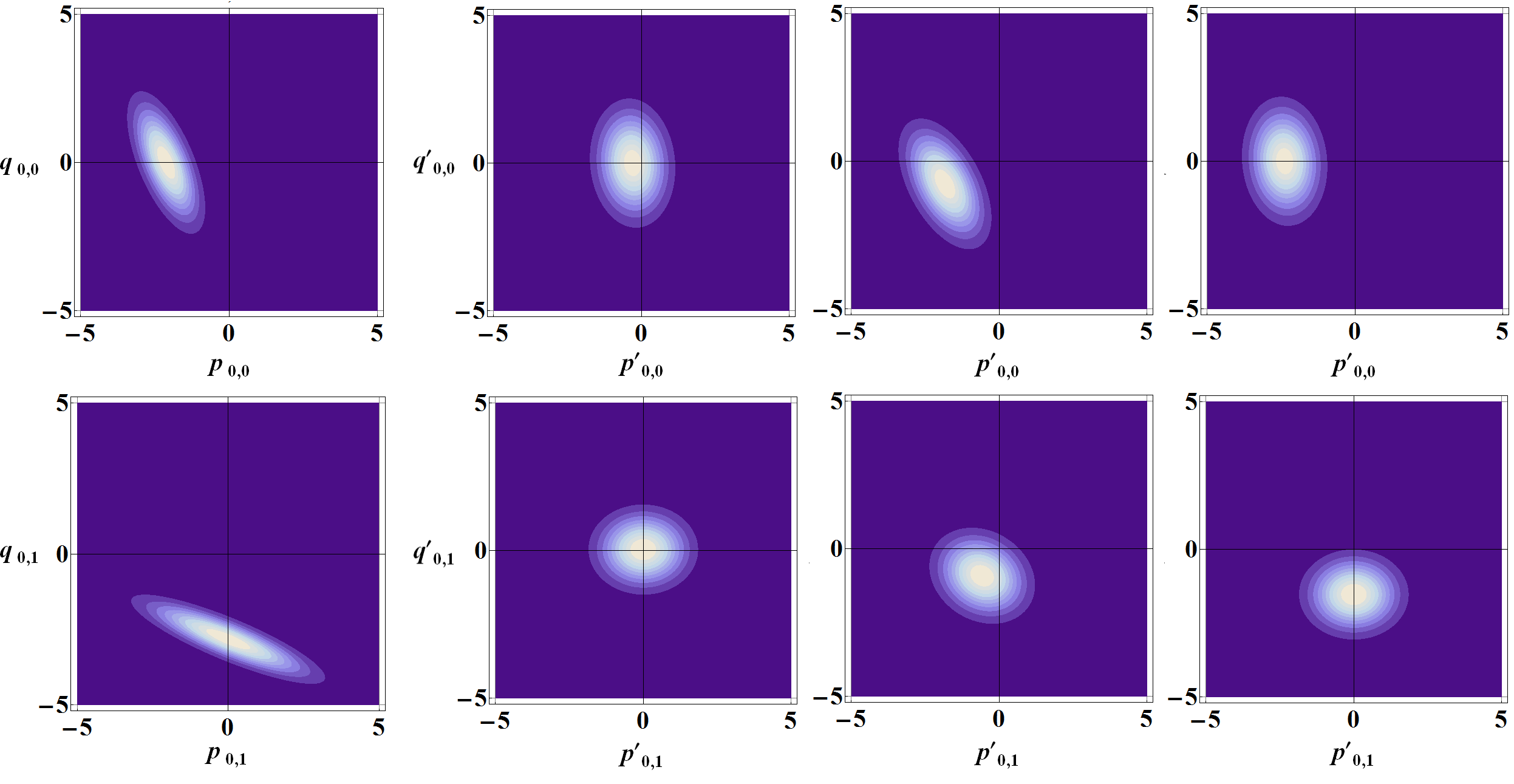}
	\caption{\label{fig:double_column_Wigner_1,0w0_narrow_contour}
		First column from the left: Input displaced-squeezed state Wigner function. Top row: LG mode $l=0, p=0$; bottom row: LG mode $l=0, p=1$.
		The input states are displaced squeezed-vacuum with displacement parameters $\alpha_0=1.5\exp(\pi i)$ and $\alpha_1=2\exp(1.5 \pi i)$, and squeezing parameters $r_0=0.5$, $\theta_0=0.25\pi$, $r_1=0.8$, $\theta_1=0.75\pi$.
		Second column:  Output quantum state Wigner function when iris is located at $z=-z_R$ (one Rayleigh range before the focus).
		Third column:  Output quantum state Wigner function when iris is located at $z=0$.
		Fourth column:  Output quantum state Wigner function when iris is located at $z=z_R$.
		 Iris radius is $w_0$. This provides evidence that moving the iris is rotating the squeezing angles via the Gouy phase. Note that the input-signal states shown in this graph are pure states, while the output-signal states in both LG modes are mixed states.
	}
\end{figure*}

Now we will show a rather surprising result, namely that the spatial mask behaves like a multi-port beam splitter with loss. To elaborate this point, let us consider the following situation. In the case of the input states of a beam splitter being two single-mode squeezed-vacuum states with identical squeezing parameters, it is well known \cite{leonhardt1997measuring} that the output state will be a two-mode squeezed state, if the beam splitter is perfectly $50:50$. Now, let us use an iris instead of a beam splitter. We put identical single-mode squeezed-vacuum states in both input LG modes ($l=0, p=0$) and ($l=0, p=1$). After the states go through the iris, we then calculate the probability of detecting $n_{0,0}$ and $n_{0,1}$ photons in the output LG modes ($l=0, p=0$) and ($l=0, p=1$), shown in Fig.~\ref{fig:Pmn_0,8339W0_1,0_1,0_0pi4_0,0pi4}. For comparison, we show the probability in the input modes as well in Fig.~\ref{fig:Pmn_InfLargeW0_1,0_1,0_0pi4_0,0pi4}. We can see in the input modes, since the quantum state is a product state of two single-mode squeezed-vacuum, the probability is non-zero only at even $n_{0,0}$ and $n_{0,1}$. If the state in the output modes is indeed a two-modes squeezed state, the probability is non-zero only at $n_{0,0}=n_{0,1}$, namely $n_{0,0}=n_{0,1}=0$, $n_{0,0}=n_{0,1}=1$, $n_{0,0}=n_{0,1}=2$, etc. However, one important visible change from the two single-mode squeezed-vacuum states to a two-mode squeezed-vacuum state is that the two-mode joint probability $P_{n_{0,0}=1,n_{0,1}=1}$ is zero in the former and non-zero in the latter \cite{Gerry_and_Knight_book}. This is indeed the case, as we can see in Fig.~\ref{fig:Pmn_InfLarge&0,8339W0_1,0_1,0_0π4_0,0π4}, which verifies our conjecture\---a hole is like a beam splitter. We can also see that the Fig.~\ref{fig:Pmn_0,8339W0_1,0_1,0_0π4_0,0π4} does not give a ideal two-mode squeezed state, this is because the iris is imbalanced (the different modes have different radial profiles) and lossy, as opposed to a perfect $50:50$ beam splitter.

\begin{figure}
	\centering
	\subfloat[\label{fig:Pmn_InfLargeW0_1,0_1,0_0pi4_0,0pi4}]{
		\includegraphics[width=0.5\columnwidth]{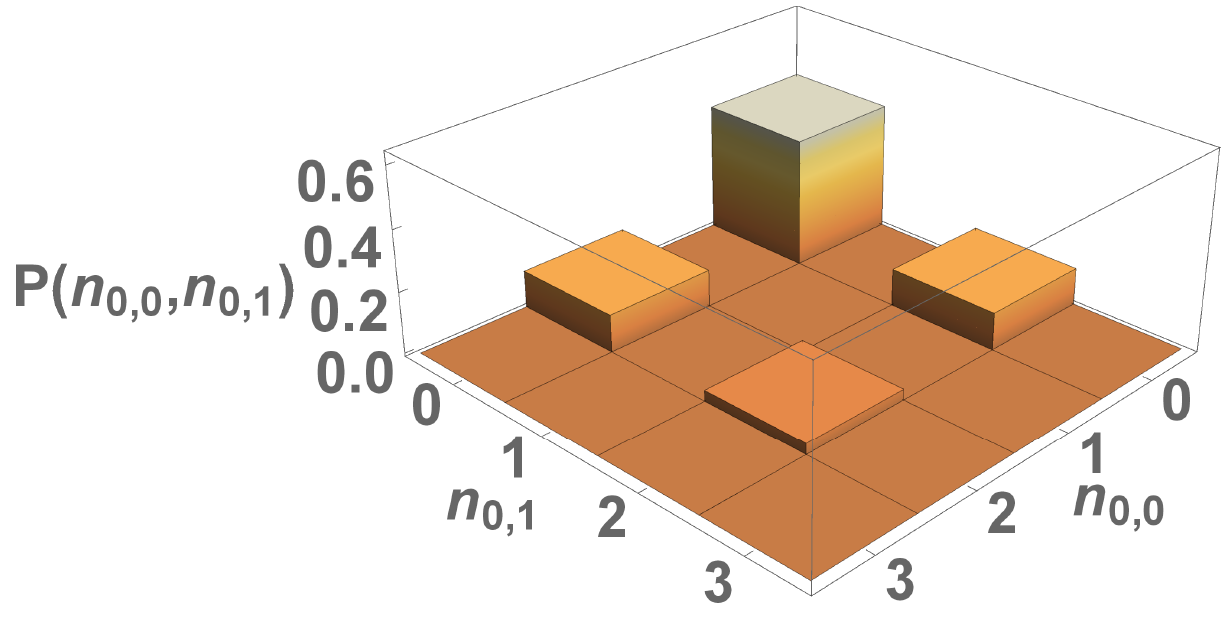}
	}
	\subfloat[\label{fig:Pmn_0,8339W0_1,0_1,0_0pi4_0,0pi4}]{
		\includegraphics[width=0.5\columnwidth]{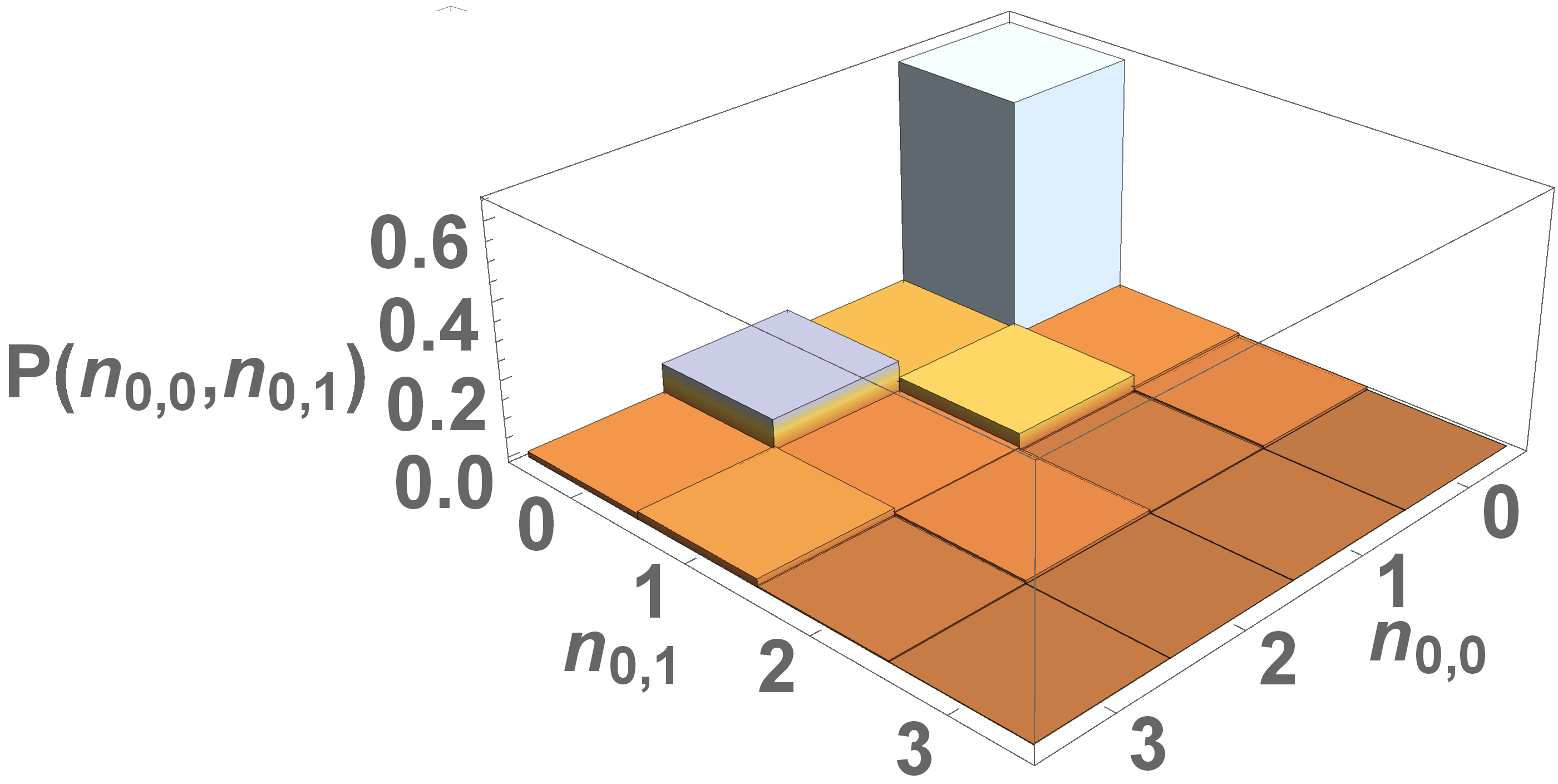}
	}
	\caption{\label{fig:Pmn_InfLarge&0,8339W0_1,0_1,0_0π4_0,0π4}
		The joint probability $P_{n_{0,0},n_{0,1}}$ vs. $n_{_{0,0}}$ vs. $n_{_{0,1}}$ in (a) input LG ($l=0, p=0$) and ($l=0, p=1$) modes, and (b) the same two LG modes in the output. The squeezing parameters of the squeezed-vacuum states in the input modes are $r_0=r_1=1, \theta_0=\theta_1=0$. The iris is placed at $z=0$ and the radius of the iris is $0.8339w_0$. Note that the non-zero probability for the one-one block provides evidence that the iris has converted the two separable squeezed-vacuum inputs into an entangled two-mode squeezed-vacuum output.}
\end{figure}

We can see how $P_{n_{0,0}=1,n_{0,1}=1}$ and $P_{n_{0,0}=3,n_{0,1}=3}$ would change with the iris size in Fig.~(\ref{fig:P11_vs_IrisSize_1,0_1,0_0pi4_0,0pi4}, \ref{fig:P33_vs_IrisSize_1,0_1,0_0pi4_0,0pi4}). Both of them reduce to zero when iris is completely closed, where the output state is reduced to vacuum. Notice $P_{n_{0,0}=1,n_{0,1}=1}$ and $P_{n_{0,0}=3,n_{0,1}=3}$ also reduce to zero in case of large iris size, where the output state is reduced to the same as the input the state (a product state of two single-mode squeezed-vacuum states). The non-zero $P_{n_{0,0}=1,n_{0,1}=1}$ and $P_{n_{0,0}=3,n_{0,1}=3}$ are what give the distinct feature of two-mode squeezing, which is most visible when the iris is neither too large nor too small, which is where the maximal interaction between LG modes ($l=0, p=0$) and ($l=0, p=1$) takes place.

\begin{figure}
	\centering
	\subfloat[\label{fig:P11_vs_IrisSize_1,0_1,0_0pi4_0,0pi4}]{
		\includegraphics[width=0.9\columnwidth]{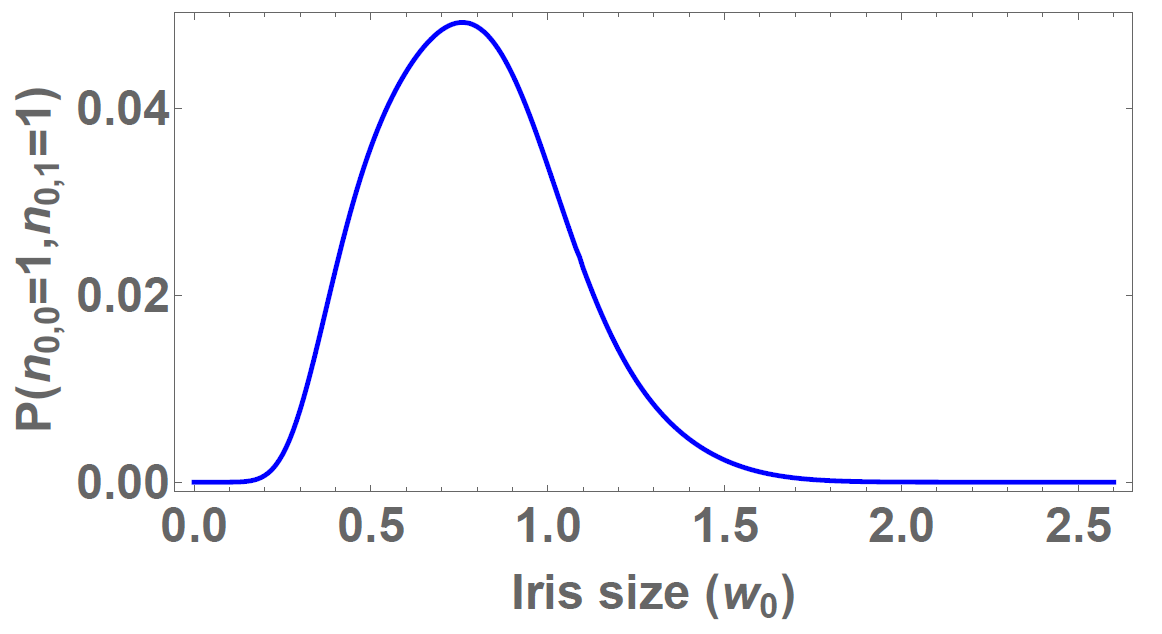}
	}
	\hspace{0mm}
	\subfloat[\label{fig:P33_vs_IrisSize_1,0_1,0_0pi4_0,0pi4}]{
		\includegraphics[width=0.9\columnwidth]{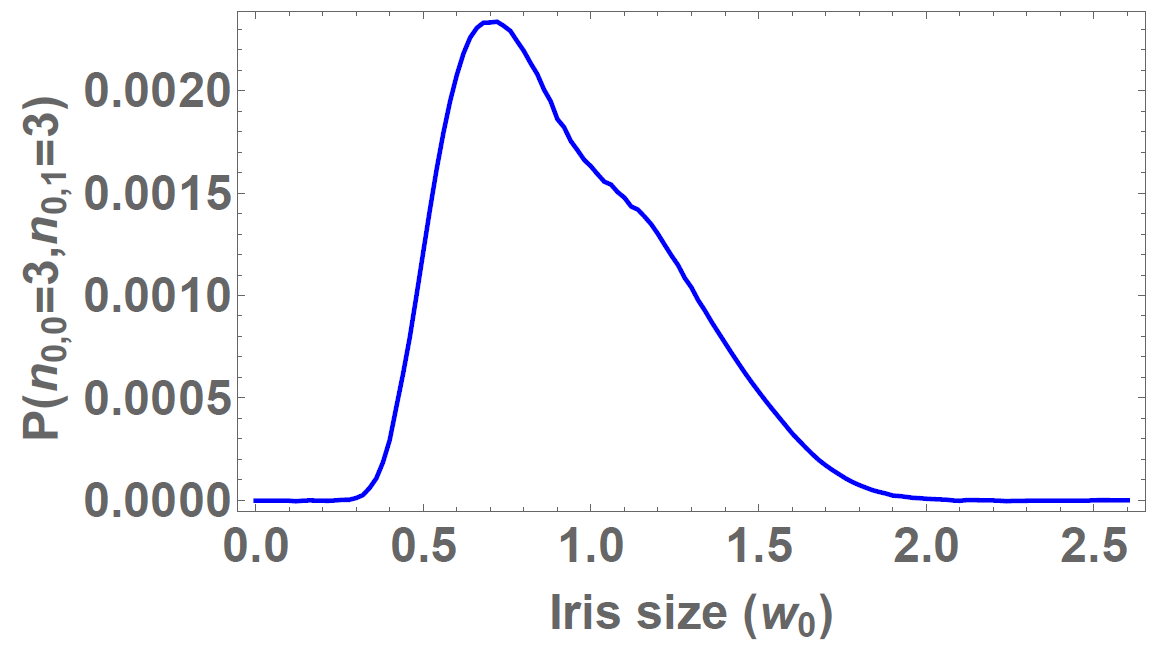}
	}
	\caption{(a) $P_{n_{0,0}=1,n_{0,1}=1}$ and (b) $P_{n_{0,0}=3,n_{0,1}=3}$ vs. iris size (scaled by $w_0$) for the output states. The squeezing parameters of the squeezed-vacuum states in the input modes are $r_0=r_1=1, \theta_0=\theta_1=0$. The iris is placed at $z=0$. If the outputs were two, separable, squeezed vacuums, then the one-one probability (a) and the three-three probability (b) would be identically zero, regardless of iris radius. The fact that both these terms are nonzero supports the idea that the output is entangled two-mode squeezed vacuum. In our beam splitter analogy the effective beam splitter is closest to $50:50$ when the iris size is about the beam waist in radius.}
\end{figure}

We can also investigate the covariance of the photon numbers in the two input modes or the two output modes, which is defined as \cite{Gerry_and_Knight_book},
\begin{equation}
\text{Cov}(n_{0,0},n_{0,1})=\braket{n_{0,0}n_{0,1}}- \braket{n_{0,0}}\braket{n_{0,1}}.
\end{equation}

For single-mode squeezed-vacuum states in the two input modes, the covariance is obviously zero since the state in each mode is independent. However, in the output modes of the iris, we should see generally non-zero covariance due to the beam-splitter-like interaction introduced by the iris, if indeed that interaction produces entangled two-mode squeezed vacuum. In this case, where we consider only LG ($l=0, p=0$) and ($l=0, p=1$) modes and two other absorption modes, the output covariance is
\begin{equation}\begin{split}\label{eqn:covariance}
&\text{Cov}(n_{0,0},n_{0,1})=C^2_{0,0,0,0}C^2_{0,0,0,1} \sinh^2r_0\cosh^2r_0\\
&+C^2_{0,0,0,1}C^2_{0,0,1,1} \sinh^2r_1\cosh^2r_1\\
&+2C_{0,0,0,0}C_{0,0,0,1}C^2_{0,0,0,1} \sinh r_0\sinh r_1\\
&\times(\sinh r_0\sinh r_1+\cosh r_0\cosh r_1\cos[4\zeta(z_0)+\theta_0-\theta_1]),
\end{split}\end{equation}
where $z_0$ is the iris position and $C_{l,l',p,p'}$'s can be calculated using Eq.~(\ref{eqn:C}). We can see in Eq.~(\ref{eqn:covariance}) the joint effect on the covariance by the Gouy phase $\zeta(z_0)$ and the squeezing angles $\theta_0$ and $\theta_1$ of the two input squeezed states. If $\theta_0$ and $\theta_1$ are different to begin with, we can counteract such difference by altering the iris position $z_0$ to change the Gouy phase. We can see how the covariance would change with iris radius in Fig.~\ref{fig:Cov_vs_IrisSize_1,0_1,0_0pi4_0,0pi4}. 

\begin{figure}
	\centering
	\includegraphics[width=1.0\columnwidth]{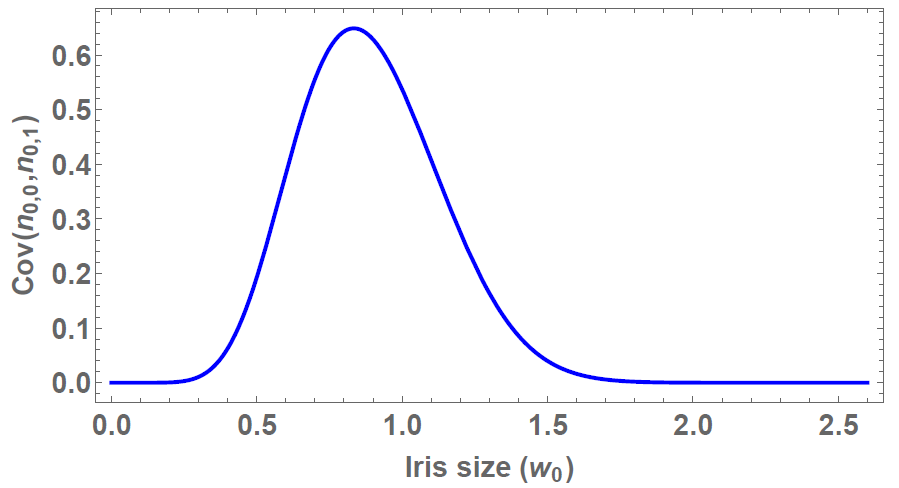}
	\caption{\label{fig:Cov_vs_IrisSize_1,0_1,0_0pi4_0,0pi4}
		Covariance of output LG modes vs. iris radius. The squeezing parameters of the two, separable, squeezed-vacuum states in the input modes are $r_0=r_1=1, \theta_0=\theta_1=0$. The iris is placed at $z=0$. We can see the covariance between the two output modes is peaked at the radius of the iris being $0.8339w_0$, which is the iris radius we use in Fig.~\ref{fig:Pmn_0,8339W0_1,0_1,0_0pi4_0,0pi4}. Also the reader might be interested to know that so long as the squeezing parameters in LG ($l=0, p=0$) and ($l=0, p=1$) modes are the same, the covariance always peaks if the iris is placed at $z=0$ with a radius of $0.8339w_0$. In our beam splitter analogy, if the outputs were again two, separable, single-mode squeezed vacuums, the covariance would be identically zero for all iris radii, which is clearly not the case. If the iris acted like a perfect $50:50$ beam splitter, the covariance would be $\frac{1}{4}\sinh^2(2r)\approx3.29$ \cite{Gerry_and_Knight_book}. However due to loss and mode mismatch it peaks here at $0.65$. Again it peaks when iris radius is about beam waist where LG ($l=0, p=0$) and ($l=0, p=1$) mode overlap is maximal.
	}
\end{figure}

To sum up, when applied to a Gaussian beam, the spatial mask behaves very much like a multi-port beam splitter with loss. If the input quantum states are displaced squeezed states, the spatial mask alters the displacement, which is a classical phenomenon; the spatial mask also alters the squeezing levels and angles, which is a non-classical phenomenon. Note that even though the input squeezed states are pure, minimum uncertainty states, the output states are generally mixed states with Wigner functions similar to displaced squeezed thermal states. The spatial mask also behaves similarly to a beam-splitter transforming a product state of two single-mode squeezed vacuums into, to an extent, an entangled two-mode squeezed state. Although this transformation is not perfect, since the spatial mask is lossy and imbalanced compared to a $50:50$ beam-splitter, there can be no doubt that even a device as simple as an iris should be treated quantum mechanically like a beam splitter.

\subsection{Example 2: Single photon in one input state and generation of number-path entanglement}

\begin{figure}
	\centering
	\includegraphics[width=1.0\columnwidth]{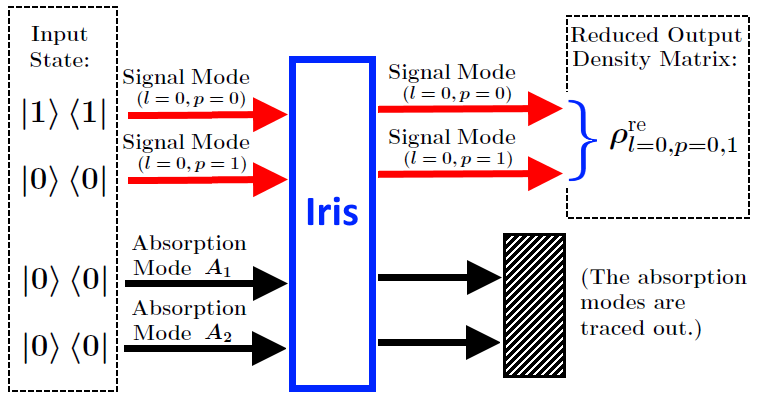}
	\caption{\label{fig:example2_diagram_shortened_twomodes_supershort_iris}
		As in the previous example, we consider two signal modes: LG mode ($l=0, p=0$) and ($l=0, p=1$), along with two absorption modes $A_1, A_2$. The input state is a product state of single photon in ($l=0, p=0$) mode and vacuum in other modes: $\ket{1}_{l=0, p=0}\otimes \ket{0}_{l=0, p=1}\otimes \ket{0}_{A_1}\otimes \ket{0}_{A_2}$. After the two output absorption modes are traced over, the reduced density matrix of the two output-signal modes, $\rho_{l=0, p=0,1}^\text{re}$, is given in Eq.~(\ref{eqn:density_matrix_example2}). We show in the Fig.~\ref{fig:Clauser_Horne_combination} that the output state has number-path entanglement, created by the iris.
	}
\end{figure}

In this example we input a single-photon state in signal mode ($l=0, p=0$) and a vacuum state in signal mode ($l=0, p=1$) as well as absorption modes $A_1, A_2$. (The input states for absorption modes are always vacuum.) Therefore the total input state in four modes is $\ket{1}_{l=0, p=0}\otimes \ket{0}_{l=0, p=1}\otimes \ket{0}_{A_1}\otimes \ket{0}_{A_2}$, shown in Fig.~\ref{fig:example2_diagram_shortened_twomodes_supershort_iris}. For the vacuum state, the corresponding Wigner function is again,
\begin{equation}\begin{split}
W_{N=0} (q,p)=\frac{1}{\pi} \exp[-(q^2+p^2)],
\end{split}\end{equation}
where $N$ is the photon number.
For the single photon state, the corresponding Wigner function is \cite{leonhardt1997measuring} 
\begin{equation}\begin{split}
W_{N=1} (q,p)=\frac{-1}{\pi} \exp[-(q^2+p^2)] L_1(2q^2+2p^2),
\end{split}\end{equation}
where $L_N$ is the $N^{\rm{th}}$ order Laguerre polynomial.
The overall Wigner function (for LG signal modes ($l=0, p=0$) and ($l=0, p=1$) and absorption modes $A_1$ and  $A_2$) is therefore,
\begin{equation}\label{eqn:example2_input_Wigner}\begin{split}
&W(q_{0,0},p_{0,0},q_{0,1},p_{0,1},q_{A1},p_{A1},q_{A2},p_{A2})\\
&=W_{N=0} (q_{0,0},p_{0,0}) W_{N=1} (q_{0,1},p_{0,1})\\
&\times W_{N=0} (q_{A1},p_{A1}) W_{N=0} (q_{A2},p_{A2}).
\end{split}\end{equation}
Using the same method in the last example we calculate the Wigner function of output modes. We then can calculate the Wigner function for either output mode, for example, LG mode ($l=0, p=0$), by tracing over the other modes:


	\begin{equation}\begin{split}
	&W_{l=0, p=0}(q_{0,0}',p_{0,0}')\\
	&=\int W(q_{0,0}',p_{0,0}',q_{0,1}',p_{0,1}',q_{A1}',p_{A1}',q_{A2}',p_{A2}')\\ &\times\text{d}q_{0,1}'\text{d}p_{0,1}'\text{d}q_{A1}'\text{d}p_{A1}'\text{d}q_{A2}'\text{d}p_{A2}'\\
	&=(1-|J_{l=0;0,0}|^2)W_{N=0} (q_{0,0}',p_{0,0}')\\
	&+|J_{l=0;0,0}|^2W_{N=1} (q_{0,0}',p_{0,0}').
	\end{split}\end{equation}
Therefore, in the output-signal LG mode ($l=0, p=0$), the reduced density operator is, 

	\begin{equation}\label{eqn:example2_reduced_density_matrix_00}\begin{split}
\rho_{{l=0, p=0}}^\text{re}=(1-|J_{l=0;0,0}|^2)\ket{0}\bra{0}+|J_{l=0;0,0}|^2\ket{1}\bra{1}.
	\end{split}\end{equation}

With a similar calculation, we find in the output-signal LG mode ($l=0, p=1$), the reduced density operator is, 

	\begin{equation}\label{eqn:example2_reduced_density_matrix_01}\begin{split}
	\rho_{{l=0, p=1}}^\text{re}=(1-|J_{l=0;0,1}|^2)\ket{0}\bra{0}+|J_{l=0;0,1}|^2\ket{1}\bra{1}.
	\end{split}\end{equation}
	
From Eqs.~(\ref{eqn:example2_reduced_density_matrix_00}, \ref{eqn:example2_reduced_density_matrix_01}) we can immediately see that if we fire a single photon in the $(l=0, p=0)$ mode and vacuum in the $(l=0, p=1)$ mode, that the photon will have a $|J_{l=0;0,0}|^2$ chance of staying in the $(l=0, p=0)$ mode at the output and a $|J_{l=0;0,1}|^2$ chance of switching to the $(l=0, p=1)$ output mode. Similarly, it is not difficult to find that if we fire a single photon in the $(l=0, p=1)$ mode and vacuum in the $(l=0, p=0)$ mode, that photon will have a $|J_{l=0;1,1}|^2$ chance of staying in the $(l=0, p=1)$ mode at the output and a $|J_{l=0;1,0}|^2|=J_{l=0;0,1}|^2$ chance of switching to the $(l=0, p=0)$ output mode. We will take another look at this result in example 3. 

However, only looking at each output mode separately does not give us the insight of correlation between modes. To achieve that we need to consider the reduced density operator for both the output-signal LG modes ($l=0, p=0$) and ($l=0, p=1$), this state is generally mixed, and perhaps more interestingly, contains number-path entanglement, which again is also created when a single photon strikes an ordinary $50:50$ beam splitter. To see this, let us examine the Wigner function in two output-signal modes:


	\begin{equation}\label{eqn:WignerSinglePhoton}\begin{split}
	&W_{l=0, p=0,1}(q_{0,0}',p_{0,0}',q_{0,1}',p_{0,1}')\\
	&=\frac{1}{\pi^2} \exp[-(q_{0,0}'^2+q_{0,0}'^2+q_{0,1}'^2+q_{0,1}'^2)]\\
	&\times \Big[(1-|J_{l=0;0,0}|^2-|J_{l=0;0,1}|^2)\\
	&+|J_{l=0;0,0}|^2 L_1(2q_{0,0}'^2+2q_{0,0}'^2)\\
	&+|J_{l=0;0,1}|^2 L_1(2q_{0,1}'^2+2q_{0,1}'^2)\\
	&+2 J_{l=0;0,0} J_{l=0;0,1}^* (p_{0,0}'-i q_{0,0}') (p_{0,1}'+i p_{0,1}')\\
	&+ 2 J_{l=0;0,0}^* J_{l=0;0,1} (p_{0,0}'+i q_{0,0}') (p_{0,1}'-i p_{0,1}')\Big].
	\end{split}\end{equation}
	
The quantum state corresponding to the Wigner Function given in Eq.~(\ref{eqn:WignerSinglePhoton}) is a mixed state of (a) the vacuum state,

\begin{equation}\begin{split}\begin{aligned}
\ket{\phi_1}=\ket{0}_{l=0, p=0}\otimes \ket{0}_{l=0, p=1},
\end{aligned}\end{split}\end{equation}
with probability of $1-|J_{l=0;0,0}|^2-|J_{l=0;0,1}|^2$, and (b) an entangled state of the form,

\begin{equation}\begin{split}\begin{aligned}
\ket{\phi_2}=\frac{J_{l=0;0,1}^*}{\sqrt{(|J_{l=0;0,0}|^2+|J_{l=0;0,1}|^2)}}\ket{0}_{l=0, p=0}\otimes \ket{1}_{l=0, p=1}\\
+\frac{J_{l=0;0,0}^*}{\sqrt{(|J_{l=0;0,0}|^2+|J_{l=0;0,1}|^2)}}\ket{1}_{l=0, p=0}\otimes \ket{0}_{l=0, p=1},
\end{aligned}\end{split}\end{equation}
with probability of $|J_{l=0;0,0}|^2+|J_{l=0;0,1}|^2$. Therefore the reduced density matrix for the output-signal LG mode ($l=0, p=0$) and ($l=0, p=1$) is,

\begin{equation}\label{eqn:density_matrix_example2}\begin{split}\begin{aligned}
\rho_{l=0, p=0,1}^\text{re}=(1-|J_{l=0;0,0}|^2-|J_{l=0;0,1}|^2)\ket{\phi_1}\bra{\phi_1}\\
+(|J_{l=0;0,0}|^2+|J_{l=0;0,1}|^2)\ket{\phi_2}\bra{\phi_2}.
\end{aligned}\end{split}\end{equation}

\begin{figure}
	[htbp]
	\centering
	\includegraphics[width=1.0\columnwidth]{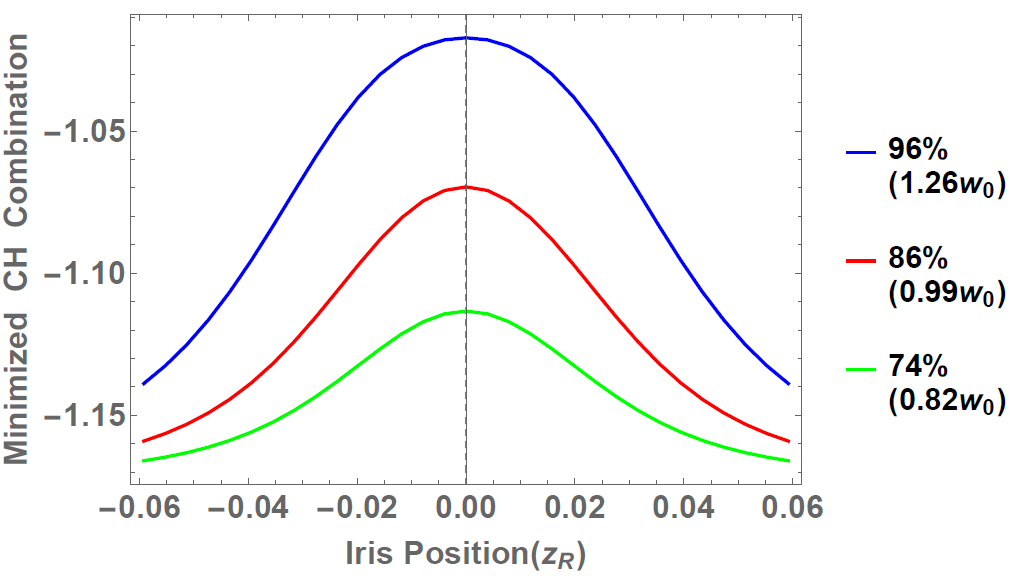}
	\caption{\label{fig:Clauser_Horne_combination}
	Minimized Clauser-Horne combination vs iris position for different sized irises. The CH combination is minimized in phase space. Notice that the minimized CH combination is below $-1$, proving the Bell inequality is violated. The more CH combination drops below $-1$, the easier the violation can be observed. In our beam splitter analogy it is well know \cite{Gerry_and_Knight_book} that a single photon incident on a $50:50$ beam splitter produces the entangled state, $\frac{1}{\sqrt{2}}(\ket{1}\ket{0}+i\ket{0}\ket{1})$.
	}
\end{figure}

One can verify this result by calculating the Wigner function of $\rho_{l=0, p=0,1}^\text{re}$ and comparing it with Eq.~(\ref{eqn:WignerSinglePhoton}). Other works have also been done to demonstrate the entanglement generation using spatial masks \cite{yu2008transforming,barbieri2004generation,torres2004quasi}. One can examine the violation of the Clauser-Horne (CH) Bell inequality \cite{clauser1974experimental,genovese2005research} of the entangled state $\ket{\phi_2}$. The more CH combination drops below $-1$, the easier the violation can be observed \cite{wildfeuer2007strong}. Therefore by plotting the minimized Clauser-Horne combination vs. iris position, shown in Fig.~\ref{fig:Clauser_Horne_combination}, we can quantitatively determine to the extent of the entanglement, which is generated by the iris, that can be observed. 


\subsection{Example 3: Single photon in each of two input states and a Hong-Ou-Mandel-like effect} 
\begin{figure}
	[htbp]
	\centering
	\includegraphics[width=1.0\columnwidth]{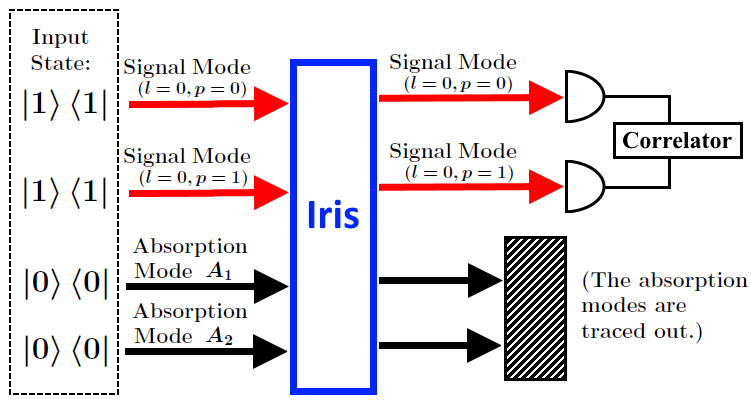}
	\caption{\label{fig:example3_diagram_shortened_twomodes_supershort_iris}
		Single-photon state in both of the two input-signal modes: LG mode ($l=0, p=0$) and ($l=0, p=1$), and vacuum state in the two input-absorption modes $A_1, A_2$. Therefore the total input state of the four modes is: $\ket{1}_{l=0, p=0}\otimes \ket{1}_{l=0, p=1}\otimes \ket{0}_{A_1}\otimes \ket{0}_{A_2}$. The coincidence probability of the two output-signal modes is given by the probability of the photon detectors receiving one photon each.
	}
\end{figure}

It is well known that when two identical photons are inputted into two separate modes of a beam splitter, photon bunching occurs. We show here the iris produces a similar effect on two LG modes. Now let us input a single photon state in mode ($l=0, p=0$) and another single photon state in mode ($l=0, p=1$). Photon detectors are used to detect the photon numbers in the two output-signal LG ($l=0, p=0$) and ($l=0, p=1$) modes, then the photon detector count signals are fed into a correlator. The setup diagram is shown in Fig.~\ref{fig:example3_diagram_shortened_twomodes_supershort_iris}. We then repeat this experiment multiple times so that we can measure the probability of detecting a single photon in each of the output-signal modes, which is refereed to as the coincidence probability. The goal is to produce a spatial mask version of Hong-Ou-Mandel (HOM) effect \cite{hong1987measurement}. Previous works have already been done to demonstrate the HOM effect in multiple spatial modes experiments \cite{karimi2014exploring,walborn2003multimode,walborn2005multimode}.

Let us first examine the case where, apart from being in different modes, the single photon state in mode ($l=0, p=0$) and the single photon state in mode ($l=0, p=1$) are completely distinguishable. This distinguishability can be caused by many reasons, such as the two photons having orthogonal polarizations or large frequency difference or large time delay when they are fired. In this case when the two photons are completely distinguishable, the two photons will not interfere and the case can be viewed simply as two independent experiments combined: (a) input one photon in mode ($l=0, p=0$) and vacuum in other modes; (b) input one photon in mode ($l=0, p=1$) and vacuum in other modes. Therefore we can simply use the analysis in example 2, and the coincidence probability is the probability that both photon staying in the same modes at the output plus the probability that both photons switching to the other modes, which is $J_{0;0,0}^2 * J_{0;1,1}^2+|J_{0;1,0}|^2 * |J_{0;0,1}|^2=J_{0;0,0}^2 J_{0;1,1}^2+|J_{0;1,0}|^4$. 

In the case the two photons are completely indistinguishable,  the coincidence probability can be calculated by the following procedure:
(a) similar to Eq.~(\ref{eqn:example2_input_Wigner}), we find the total input state Wigner function: 

\begin{equation}\begin{split}
&W(q_{0,0},p_{0,0},q_{0,1},p_{0,1},q_{A1},p_{A1},q_{A2},p_{A2})\\
&=W_{N=1} (q_{0,0},p_{0,0})  W_{N=1} (q_{0,1},p_{0,1})\\
&\times W_{N=0} (q_{A1},p_{A1}) W_{N=0} (q_{A2},p_{A2});
\end{split}\end{equation}
(b) and we find the total output state Wigner function $W(q_{0,0}',p_{0,0}',q_{0,1}',p_{0,1}',q_{A1}',p_{A1}',q_{A2}',p_{A2}')$ using the transformation described in Eq.~(\ref{eqn:transformQ}) and (\ref{eqn:transformP});
(c) and we find the reduced density matrix for the output-signal modes by tracing over the absorption modes:

	\begin{equation}\begin{split}
	&W_{l=0, p=0; l=0, p=1}(q_{0,0}',p_{0,0}',q_{0,1}',p_{0,1}')\\
	&=\int W(q_{0,0}',p_{0,0}',q_{0,1}',p_{0,1}',q_{A1}',p_{A1}',q_{A2}',p_{A2}')\\ 
	&\times\text{d}q_{A1}'\text{d}p_{A1}'\text{d}q_{A2}'\text{d}p_{A2}' ;
	\end{split}\end{equation}
(d) and we find the Wigner function for state of single photon in each output-signal mode, which is the state when coincidence is detected $\rho_{\text{coincidence}}=\ket{1}_{l=0, p=0}\bra{1}\otimes \ket{1}_{l=0, p=1}\bra{1}$:

\begin{equation}\begin{split}
&W_{\text{coincidence}}(q_{0,0}',p_{0,0}',q_{0,1}',p_{0,1}')\\
&=W_{N=1} (q_{0,0}',p_{0,0}') \times W_{N=1} (q_{0,1}',p_{0,1}');
\end{split}\end{equation}
(e) and we find the coincidence probability by projecting the output reduced density matrix onto the state $\rho_{\text{coincidence}}$ and calculating the trace:

\begin{equation}\begin{split}
&P_{\text{coincidence}}\\
&=\rm{tr}[\rho_{l=0, p=0,1}^\text{re} \ket{1}_{l=0, p=0}\bra{1}\otimes \ket{1}_{l=0, p=1}\bra{1}]\\
&=\int W_{l=0, p=0; l=0, p=1} W_{\text{coincidence}} \text{d}q_{0,0}'\text{d}p_{0,0}'\text{d}q_{0,1}'\text{d}p_{0,1}'\\
&=J_{0;0,0}^2 J_{0;1,1}^2+|J_{0;1,0}|^4+2 J_{0;0,0} J_{0;1,1} |J_{0;1,0}|^2.
\end{split}\end{equation}

\begin{figure}
	\centering
	\includegraphics[width=1.0\columnwidth]{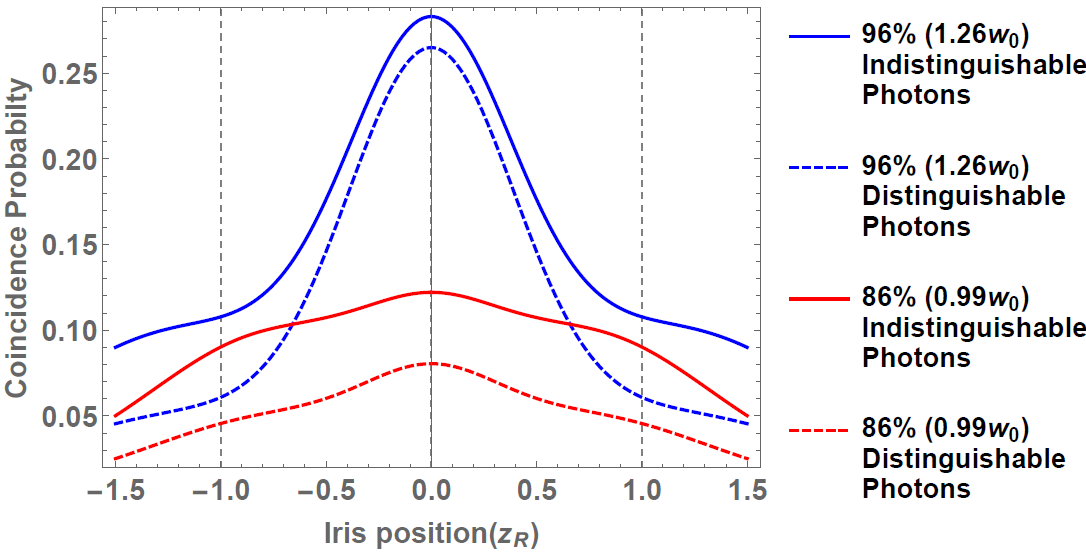}
	\caption{\label{fig:coincidence_prob_vs_iris_posi}
		Coincidence probability vs. iris position. Lines of different colors represent different iris sizes, denoted by the percentage of transmitted beam intensity through the iris at focus point (relative to full beam intensity) as well as iris radius (relative to $w_0$). The solid lines represent the indistinguishable photons inputed in two signal modes as opposed to the dashed lines  representing the distinguishable photons. We can see that for the iris of same size and placed at the same position at the beam axis, inputting indistinguishable photons always leads to a higher coincidence probability than distinguishable photons. This we call a Hong-Ou-Mandel bump, and it is a hallmark of two-photon interference.
	}
\end{figure}

Therefore, the coincidence probability for indistinguishable photons $J_{0;0,0}^2 J_{0;1,1}^2+|J_{0;1,0}|^4+2 J_{0;0,0} J_{0;1,1} |J_{0;1,0}|^2$, is greater than coincidence probability for distinguishable photons $J_{0;0,0}^2 J_{0;1,1}^2+|J_{0;1,0}|^4$, since $J_{0;0,0}, J_{0;1,1}, |J_{0;1,0}|^2 \geq 0$, as shown in Fig.~\ref{fig:coincidence_prob_vs_iris_posi}.

Therefore in the iris version of HOM effect, when photons in the two input-signal modes are made to be indistinguishable, the coincidence probability rises. This is in contrast with the beam splitter version of HOM effect, in which the coincidence probability falls when photons in two input ports are made to be indistinguishable. In short, the iris produces a HOM ``bump'' while the beam splitter produces a HOM ``dip''.

The physical interpretation of the ``bump'' is that, while a beam splitter introduces a $\pi$ phase shift on the reflected beam, the spatial mask in our case does not produce any phase shift when the same two LG modes in the incident and diffracted beams are considered. This fermion-like anti-bunching behavior has been more thoroughly investigated in Ref.~\cite{matthews2013observing}.
\\


\section{Conclusion} 
We have analyzed the Gaussian beams both classically and quantum mechanically. We have developed a clear method to calculate the interaction between quantum states in various Gaussian modes. While we focus on the mode interactions introduced by an iris, it is straightforward to extend our method to other spatial masks or other optical devices. The framework we established allows us to analyze arbitrarily many Gaussian modes including various orbital angular momentum LG modes as well as HG modes. We verified our theory via our experiment, in which we generated squeezed states in various LG modes and found that the experimental data agreed with our numerical simulation. We finally gave three examples to show some interesting phenomenon that can be easily tested in future experiments. These examples are displaced (and non-displaced) squeezed-vacuum input states, along with single and double photon input states. These examples predict that the diffraction process gives rise to photon-number entanglement and a Hong-Ou-Mandel-like effect, which implies the spatial mask behaves in ways similar to an ordinary beam splitter. As we pointed out, the purpose of this work is to setup a general method for analyzing quantum states interaction between Gaussian modes, which can be useful in many ways, including: creating a specific quantum state in higher-order modes from lower-order modes, creating entanglement between modes, optimizing overall squeezing, designing specific interaction between different OAM modes, etc.

\section{Acknowledgments} 
This research was supported by Air Force Office of Scientific Research grant FA9550-13-1-0098. In
addition, Z. X., R. N. L. and J. P. D. would also like to acknowledge
support from the Army Research Office, the National Science Foundation, and the Northrop Grumman Corporation.

\bibliography{A_general_diffraction_theory_for_multimode_quantum_states_of_light_arxiv_03102017}

\end{document}